\documentclass[aps,prl,amsmath,amssymb,reprint,groupedaddress]{revtex4-2}
\usepackage[utf8]{inputenc}
\usepackage[english]{babel} 
\usepackage[T1]{fontenc}
\usepackage{graphicx}
\usepackage{placeins}
\usepackage{amsmath,amssymb}
\usepackage{cancel}
\usepackage[colorlinks=true,citecolor=red]{hyperref}		% adds hyperlinks to .pdf file
\usepackage{natbib}
\usepackage{xcolor}
\usepackage{siunitx}
\usepackage{braket}

\newcommand{\rb}{$^{87}$Rb}

\begin{document}

\title{Multimode feedback cooling of the collective modes of a Bose-Einstein condensate}
\author{Ryan J. Thomas}\email{ryan.thomas@anu.edu.au}
\author{Jordan A. McMahon}\altaffiliation[Current address: ]{Q-CTRL, 93 Shepherd St, Chippendale, NSW 2008, Australia.}
\author{Zain Mehdi}
\author{Stuart S. Szigeti}
\author{Simon A. Haine}
\author{Samuel Legge}
\author{John D. Close}
\author{Joseph J. Hope}
\affiliation{Department of Quantum Science and Technology, The Australian National University, Canberra, Australia}

\begin{abstract}

We experimentally demonstrate cavity-free feedback cooling of the three lowest-lying collective modes of a Bose-Einstein condensate in a prolate harmonic trap.  Using shadowgraph imaging as an \textit{in situ} probe of the atomic density, we measure the time-dependent centers of mass and widths of the condensate in two dimensions and use these measurements to damp oscillations in the two visible dipole modes and the low-frequency quadrupole mode.  By inducing oscillations in the condensate, we show that we can efficiently damp the dipole modes to a final mean phonon occupancy per atom of $<1$.

\end{abstract}

\maketitle

\emph{Introduction.} With few exceptions, degenerate quantum gases are produced using essentially the same method as the first demonstrations of Bose-Einstein condensation \cite{anderson_observation_1995,davis_bose-einstein_1995,bradley_evidence_1995}, where an initial laser cooling stage is followed by evaporative cooling in a conservative trap to reach degeneracy.  Evaporative cooling is extremely lossy, with typically $99.9\%$ of the atoms ejected from the trap, which limits the number of degenerate atoms to $<\!10^7$ \cite{andrews_direct_1996,raman_evidence_1999} even with machine optimization of the evaporation sequence \cite{wigley_fast_2016, barker_applying_2020, vendeiro_machine-learning-accelerated_2022}.  Although it is possible to directly laser cool atoms to degeneracy \cite{stellmer_laser_2013,urvoy_direct_2019,chen_continuous_2022, vendeiro_machine-learning-accelerated_2022}, these techniques require tightly confining traps where two- and three-body loss limits the total number of atoms.

Feedback cooling, which uses closed-loop control to damp excitations, is a powerful tool for cooling quantum systems \cite{zhang_quantum_2017} and has been applied to trapped ions \cite{bushev_feedback_2006}, single atoms \cite{fischer_feedback_2002,kubanek_photon-by-photon_2009}, optical cavities \cite{cohadon_cooling_1999}, and optomechanical systems \cite{ashkin_feedback_1977, gieseler_subkelvin_2012, li_millikelvin_2011, wilson_measurement-based_2015, vovrosh_parametric_2017, tebbenjohanns_cold_2019, tebbenjohanns_quantum_2021, magrini_real-time_2021}.  In these experiments, a single degree of freedom is cooled by the addition of a time-dependent external potential which is changed according to real-time measurements of the system's state.  Ultracold atomic gases, in contrast, can comprise anywhere between $10^4$ and $10^{10}$ atoms, enormously increasing the number of degrees of freedom that need to be cooled.  Although theoretical investigations of feedback cooling have shown that it is possible to cool Bose gases to degeneracy even in the presence of measurement backaction and heating from spontaneous emission \cite{haine_control_2004, szigeti_continuous_2009, szigeti_feedback_2010, hush_controlling_2013, mehdi_fundamental_2024,zhu_simulating_2025}, implementing closed-loop control on the required timescales to remove energy from thermal fluctuations is extremely challenging.

Instead of this ``top-down'' approach to feedback cooling, where a non-degenerate gas is cooled through the transition point, Bose-Einstein condensates (BECs) can be subject to ``bottom-up'' feedback cooling \cite{mehdi_multi-mode_2025}.  In this approach, a finite-temperature BEC is formed using evaporative cooling, and residual energy is removed through a feedback cooling protocol that targets low-energy modes; in principle, higher-energy modes are damped by thermalization \cite{szigeti_feedback_2010,mehdi_fundamental_2024}.  Direct feedback-cooling of these low-energy collective modes offers clear benefits, as these modes can be excited by evaporative cooling itself \cite{kuwamoto_collective_2012} or when the trap deliberately or inadvertently moves relative to the BEC \cite{gaaloul_space-based_2022}, and they are inefficiently cooled by evaporative cooling which must selectively target high-energy excitations \cite{olf_thermometry_2015}. Bottom-up feedback cooling has significantly reduced spatiotemporal resolution requirements compared to the top-down approach---and is therefore simpler to implement---since low-energy excitations oscillate at frequencies on the order of the trap frequency and have long spatial wavelengths on the order of the sample size. Nevertheless, lessons learned from investigations into bottom-up feedback cooling directly inform top-down feedback cooling; for example, a BEC is more sensitive to measurement noise in the feedback cooling loop than a thermal gas, so the bottom-up approach is a crucial test of the feasibility of using feedback to cool atomic gases to degeneracy. Although feedback cooling of an ultracold atomic gas has been demonstrated \cite{morrow_feedback_2002}, that experiment's use of stimulated Bragg scattering precluded control over more than a single degree of freedom: to our knowledge, no previous experiments have demonstrated multi-mode feedback cooling of atomic gases.

In this Letter, we report a demonstration of feedback cooling of the three lowest-energy collective modes of a harmonically trapped BEC.  Using nondestructive shadowgraph imaging, and in the absence of an optical cavity, we measure the two-dimensional spatial density of the trapped BEC from which we can extract the amplitudes of the visible dipole (center-of-mass oscillations) and quadrupole modes.  We modify the potential energy in real time by modulating the position and trapping frequencies of our optical potential based on information obtained from the measurement time series and show that we can cool all three modes of the BEC, with the dipole modes having a final mean phonon occupancy per atom of less than unity.  Our results establish bottom-up feedback cooling as an immediately feasible technique for the control and cooling of quantum gases.

\begin{figure}[t]
	\centering
	\includegraphics[width=\columnwidth]{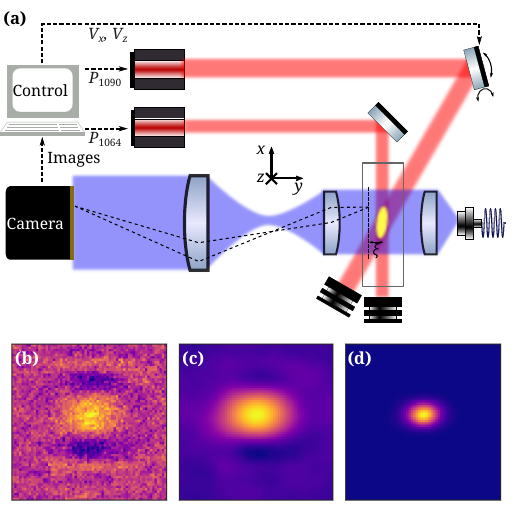}
	\caption{\textbf{(a)} Simplified optical setup of the feedback cooling experiment.  A BEC (yellow) is formed at the intersection of two far off-resonant laser beams and is imaged in trap using a CMOS camera and off-resonant shadowgraph imaging, where the imaging system is focused a distance $\xi$ from the sample.  Images are retrieved by the processor, from which the BEC's real-time center of mass and width in the $x$ and $z$ directions are determined. These extracted moments are used to compute actuator values that drive the laser powers and angles of the dipole trap, which control the motion of the BEC.  \textbf{(b) - (d)} Example false-colour images showing image processing steps.  \textbf{(b)} Raw shadowgraph image.  \textbf{(c)} Density estimate $\hat{\rho}(x,z)$.  \textbf{(d)} Density estimate after nonlinear filter $[\hat{\rho}(x,z)]^6$.}
	\label{fg:apparatus}
\end{figure}

\emph{Experiment.} The experimental apparatus has been described previously \cite{kuhn_bose-condensed_2014,everitt_observation_2017}, and a simplified schematic is shown in Fig.~\ref{fg:apparatus}(a).  We trap a sample of \rb{} atoms in $\ket{F = 1,m_F = -1}$ in an optical dipole trap consisting of two laser beams with wavelengths $1064$ and \SI{1090}{\nano\meter} and which intersect at an angle of $\mathord{\approx}30^\circ$.  The trapped sample is evaporatively cooled by reducing the power in both lasers, producing a BEC of approximately $4\times 10^5$ atoms.  We then adiabatically increase the laser power over \SI{200}{\milli\second} to increase the trap depth in preparation for feedback cooling which results in a final condensate fraction of $50\%$ with trapping frequencies $(\omega_x,\omega_y,\omega_z) = 2\pi\times(20.3,85.6,70.3)$ \si{\hertz}.  We implement feedback cooling through coupled control of the laser powers and the $(x,z)$ location at which the \SI{1090}{\nano\meter} laser intersects the \SI{1064}{\nano\meter} laser through piezoelectric mirror mounts.  Together, these four actuators give us control over the harmonic trap's centers and frequencies in two dimensions (see Supplemental Material \cite{supp}).

We measure the \textit{in situ} integrated column density of the atomic sample in the $x$-$z$ plane using off-resonant shadowgraph imaging \cite{bradley_evidence_1995,andrews_direct_1996,wigley_non-destructive_2016}. This technique forms an image of the near-field Fresnel diffraction pattern on the camera which, in the limit of a flat illuminating field, negligible absorption, and a small phase shift, is approximately the transverse Laplacian in the $x$-$z$ plane of the object's column-integrated phase profile~\cite{supp} [Fig.~\ref{fg:apparatus}(b)].  The spatial density, which is proportional to the imprinted phase, can thus be efficiently computed through a Fourier-domain filter of the measured intensity pattern.

Our imaging laser is red-detuned by $\mathord{\approx}900$ linewidths, and the image of the Fresnel diffraction pattern is formed on a CMOS sensor with a \SI{5.5}{\micro\meter} resolution.  We pulse the imaging light on for \SI{20}{\micro\second} every \SI{1}{\milli\second} at an intensity of $\approx\!\SI{70}{\watt/\meter^2}$, and we measure a heating rate of \SI{100}{\pico\kelvin} per image.  This rate is about twice as large as theoretically expected from spontaneous emission alone, and the excess is likely due to the measurement beam forming stray optical lattices via reflections from the vacuum cell windows~\cite{altuntas_weak-measurement-induced_2023}.  We record and process images in real time by restricting the camera to read only the small subset of the total image where the atomic signal is non-zero. We estimate the two-dimensional atomic density $\hat{\rho}(x,z)$ by subtracting a reference image to eliminate spurious fringes [Fig.~\ref{fg:apparatus}(b)] and applying a regularized inverse Laplacian operator $\nabla^{-2}$ in the Fourier domain [Fig.~\ref{fg:apparatus}(c)]~\cite{supp}.  The inverse Laplacian operator attenuates high spatial frequency noise at the cost of amplifying low-frequency noise, which is dominated by imperfect subtraction of spurious fringes.  

The BEC's lowest-order collective modes are the dipole and quadrupole modes, which change the position and size of the sample.  We therefore measure the cloud's centers of mass $\hat{x}(t)$ and $\hat{z}(t)$ and the widths $\hat{w}_x(t)$ and $\hat{w}_z(t)$ by computing the first moment and second central moment for each dimension.  Moment calculations are sensitive to outliers, so we apply a nonlinear filter by first computing $[\hat{\rho}(x,z)]^6$ before calculating the moments, which improves the signal-to-noise ratio (SNR) in regions where it is greater than $1$ \cite{supp}.  Although we compute $\hat{w}_z(t)$, we do not use it for feedback because the equilibrium size of the BEC in the $z$ direction ($\mathord{\approx}\SI{5}{\micro\meter}$) is on the order of the imaging system's resolution ($\SI{5.5}{\micro\meter}$), and so we observe only weak temporal variation in $\hat{w}_z(t)$. 

Each collective mode behaves as an independent harmonic oscillator, so we implement a purely derivative control law to increase the effective damping rate.  We compute the derivatives as finite differences between successive measurements, which together form the vector error signal for the system state \cite{supp}.  Actuator values are then computed through matrix multiplication of the feedback gain matrix and the vector error signal.  To control the dipole modes, we adjust the trap position by changing the \SI{1090}{\nano\meter} laser angle through piezoelectric actuators, and to control $\hat{w}_x(t)$ we change the \SI{1090}{\nano\meter} laser power which changes $\omega_z$.  Gravitational sag of the trap in the $z$ direction creates unwanted coupling between the \SI{1090}{\nano\meter} laser power and the vertical position, resulting in large vertical oscillations when the power is changed.  We mitigate this effect by simultaneously changing the \SI{1064}{\nano\meter} laser power but in the opposite direction so that the gravitational sag remains approximately constant.  The feedback gains relating $\hat{w}_x(t)$ to the laser powers are then chosen to minimize the perturbation to $\hat{z}(t)$.  

For simplicity, we implement the entire control system using LabVIEW software on a standard desktop computer.  Although convenient, there is a delay of \SI{960}{\micro\second} between when the camera starts the exposure and when the control hardware responds, and nearly all of this delay is communication overhead between the computer, camera, and control hardware.  This delay limits our control loop to frequencies much less than \SI{165}{\hertz}, and thus control over $\hat{z}(t)$, which has a frequency of \SI{70}{\hertz}, is more limited than control over the horizontal modes.  As demonstrated theoretically \cite{mehdi_multi-mode_2025}, we have found that including a digital first-order low-pass filter on the measurements $\hat{x}(t)$ and $\hat{w}_x(t)$, with cutoff frequencies $60$ and \SI{100}{\hertz}, respectively, improves feedback performance for those quantities, and including a similar filter on the output powers but with a cut-off frequency of \SI{100}{\hertz} reduces unwanted excitation of the vertical dipole mode.  We do not include filters on the measurement of $\hat{z}(t)$ because it would introduce additional delay causing further degradation of feedback.

\emph{Temporal behavior.} After evaporative cooling, the BEC's small residual motion in the targeted collective modes is below our measurement noise, so to demonstrate feedback cooling we deliberately induce large-amplitude oscillations in these modes.  In the first experiment, we primarily excite the dipole modes by suddenly displacing the horizontal trap center and simultaneously increasing the \SI{1090}{\nano\meter} laser power, the latter of which also generates a small quadrupole excitation.  In a second experiment, we primarily excite the quadrupole mode by sinusoidally modulating the \SI{1090}{\nano\meter} laser power close to the mode frequency for four periods; this method allows us to drive large amplitude excitations without driving large excitations in the vertical dipole mode which can cause the BEC to leave the trap.

\begin{figure}[t]
	\centering
	\includegraphics[width=\columnwidth]{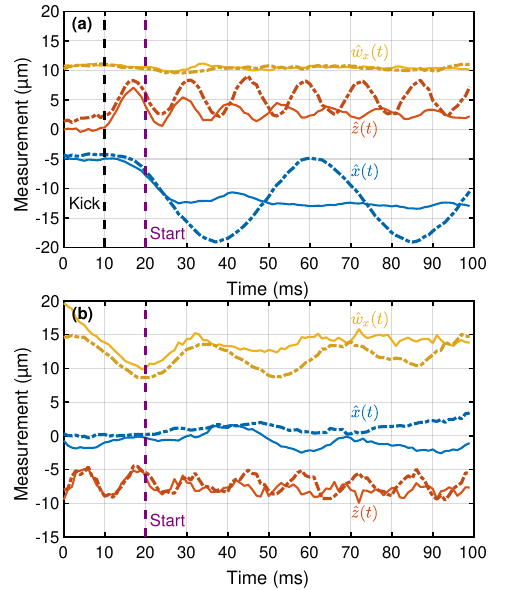}
	\caption{Measured signals $\hat{x}(t)$, $\hat{z}(t)$, and $\hat{w}_x(t)$ (colored blue, red, and yellow, respectively) after real-time filtering.  Solid and dashed lines indicate signals with feedback enabled and disabled, respectively. The vertical purple dashed line at \SI{20}{\milli\second} shows the time at which feedback is engaged for those signals where it is enabled. \textbf{(a)} Excitation of dipole modes in the horizontal and vertical directions.  The vertical black dashed line at \SI{10}{\milli\second} indicates the ``kick'' given to the atoms by shifting the trap's horizontal position \SI{-8}{\micro\meter} and increasing the \SI{1090}{\nano\meter} laser power by \SI{0.3}{\watt}, causing large oscillations in $\hat{x}(t)$ and $\hat{z}(t)$.  \textbf{(b)} Excitation of the axial quadrupole mode.  The periodic driving used to excite the quadrupole mode happens prior to the start of imaging.}
	\label{fg:temporal-results}
\end{figure}  

In Fig.~\ref{fg:temporal-results}(a), we show the real-time measurement records from exciting the dipole modes by suddenly changing the trap equilibrium position at $t=\SI{10}{\milli\second}$. When enabled, feedback is applied from $t = \SI{20}{\milli\second}$.  Damping of the horizontal dipole mode $\hat{x}(t)$ is the most pronounced, with the mode reaching equilibrium in only \SI{10}{\milli\second}, corresponding to near-critical damping.  In contrast, the vertical dipole mode $\hat{z}(t)$ takes the entire record to reach an amplitude near its noise floor, and the reduced damping is a consequence of the feedback-loop delay which limits the maximum possible feedback gain.  Although the quadrupole mode experiences only a minor excitation, some damping of the mode can be seen.

In contrast, Fig.~\ref{fg:temporal-results}(b) shows the temporal behavior of the three collective modes after exciting the axial quadrupole mode by modulating the trap frequency. When enabled, feedback is applied from $t = \SI{20}{\milli\second}$.  As the quadrupole frequency is $\mathord{\approx}\sqrt{5/2}\omega_x = 2\pi\times\SI{32}{\hertz}$, it is far below the frequency limit of the control loop, and we can use high feedback gains to induce rapid damping of the mode.  Residual coupling of the laser powers to the vertical mode manifests as additional, higher-frequency noise in $\hat{z}(t)$ when feedback is enabled, since the feedback loop attempts to control the quadrupole mode through laser power, and it inadvertently excites the high-$Q$ vertical dipole mode.

\emph{Degree of cooling.} We investigate the effectiveness of feedback cooling by exciting the sample, alternately applying or not applying feedback, waiting a period of time, and finally taking an absorption image after \SI{20}{\milli\second} time of flight.  We perform $1500$ experiments for the two excitation strategies to obtain accurate statistics.  When exciting the dipole modes, we wait a random time in the interval $[0,150]$ \si{\milli\second} before releasing the atoms from the trap, which randomizes the phase of the dipole oscillations at the release time.  When exciting the quadrupole mode, we wait \SI{500}{\milli\second} after the last shadowgraph image to ensure that the sample has thermalized.  

\begin{figure}[t]
	\centering
	\includegraphics[width=\columnwidth]{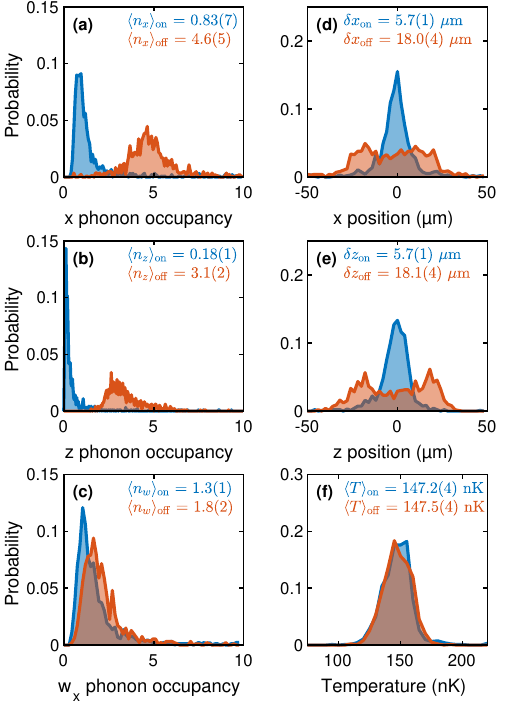}
	\caption{\textbf{(a)} - \textbf{(c)} Distribution of in-trap phonon occupancies for the horizontal dipole, vertical dipole, and axial quadrupole modes when the dipole modes are excited.  \textbf{(d) - (f)} Distributions of out-of-trap measurements using absorption imaging after a time-of-flight of \SI{20}{\milli\second}.  Blue and red shaded areas are when feedback is enabled and disabled, respectively.  Reported mean phonon occupancies are true phonon values; see text for details.}
	\label{fg:dipole-results}
\end{figure}

We quantify cooling effectiveness using both in-trap and out-of-trap measurements.  For the in-trap characterization, we fit a model of the shadowgraph image to the raw shadowgraph image data and extract estimates of the position in two dimensions and the width in the axial direction \cite{supp}.  For each temporal record, we compute the measured phonon occupancy per atom $n_{i, \rm meas}$ of the $i$th mode:
\begin{equation}
	n_{i,{\rm meas}} = \frac{1}{2}\frac{\mathrm{Var}(r_i - r_{i,\rm trap})}{a_{{\rm ho},i}^2} + \frac{1}{2}\frac{\mathrm{Var}(\dot{r}_i)}{a_{{\rm ho},i}^2\omega_i^2},
\end{equation}
where $\textrm{Var}(\cdot)$ is the variance of the argument over the last period of the $i$th mode, $\omega_i$ is the frequency of that mode, and $a_{{\rm ho},i} = \sqrt{\hbar/m\omega_i}$.  Here, $r_i$ and $\dot{r}_i$ are the measured position and velocity of the mode, with $\dot{r}_i$ computed using finite differences, and for the dipole modes $r_{i,\rm trap}$ is the equilibrium position of the optical trap.  Post-processing measurement noise $\delta r_i$ biases the measured mean phonon number $n_{i,\rm meas}$, so that the ``true'' value is
\begin{equation}
	n_{i,\rm true}= n_{i,\rm meas} - \frac{1}{2}m\omega_i^2\mathrm{Var}(\delta r_i)\left[1 + \frac{2}{\omega_i^2\tau^2}\right],
\end{equation} 
for sampling time $\tau = \SI{1}{\milli\second}$ \cite{supp}, and we find phonon biases $n_{i,\rm meas} - n_{i,\rm true}$ for the $x$, $z$, and $w_x$ modes to be $0.156$, $0.046$, and $0.099$, respectively.  The out-of-trap measurements are formed from the absorption image taken for each run of the experiment.  Using a two-dimensional fit to the bi-modal density distribution, we extract the $(x,z)$ center of the atomic sample and the temperature $T$ of the thermal fraction.

Figure~\ref{fg:dipole-results} shows the results of our analysis for excitation of the dipole modes.  Although we apply a deterministic ``kick'' to the atoms, changes in the initial conditions and actuator response over time mean that the actual excitations have a broad distribution of phonon occupancies [Figs.~\ref{fg:dipole-results}(a-c)].  When engaged, we achieve ground-state cooling of both dipole modes where feedback reduces the mean phonon occupancy to $n_{i,\rm true} < 1$.  Feedback is less effective at cooling the quadrupole mode, due to both weak excitation and limited SNR on the measurement of $w_x$, which is partly a consequence of using the nonlinear filter to enhance the measurement of the dipole modes.  The out-of-trap measurements in Figs. \ref{fg:dipole-results}(d-e) confirm the reduction of the in-trap dipole oscillations, and the phonon occupancies when feedback is engaged agree well with the residual position uncertainty after time of flight; for the $x$ and $z$ modes, we predict residual position uncertainties after a \SI{20}{\milli\second} time-of-flight of \SI{5.8}{\micro\meter} and \SI{4.9}{\micro\meter}, respectively. The out-of-trap measurement in Fig.~\ref{fg:dipole-results}(f) also shows that feedback control does not affect the temperature of the sample, although the measurements---which are present both when feedback is on and off---do cause heating of the sample on the order of \SI{10}{\nano\kelvin}.  Since in purely harmonic traps the dipole modes are decoupled from all other collective modes \cite{lee_non-equilibrium_2016}, cooling of these modes will not affect the sample temperature.

\begin{figure}[t]
	\centering
	\includegraphics[width=\columnwidth]{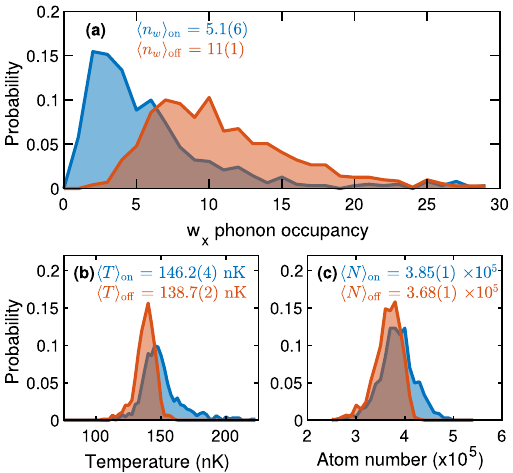}
	\caption{  \textbf{(a)} Distribution of in-trap quadrupole phonon occupancy after excitation with a sinusoidal drive.  Reported mean values are true values after correcting for measurement noise.  \textbf{(b)} Distribution of the thermal fraction temperature after \SI{20}{\milli\second} time of flight.  \textbf{(c)} Distribution of the number of atoms as measured with absorption imaging.  In all panels, blue and red shaded areas are when feedback is enabled and disabled, respectively.}
	\label{fg:quadrupole-results}
\end{figure}

\emph{Quadrupole excitation.} Unlike dipole modes, quadrupole modes are coupled to other even-symmetry collective modes \cite{haine_control_2004} which causes damping of oscillations [Fig.~\ref{fg:temporal-results}(b)] and thus thermalization of the excited mode with the rest of the condensate.  In Fig.~\ref{fg:quadrupole-results}, we show the effect of feedback cooling on the quadrupole mode when it is strongly excited by sinusoidal driving.  Similarly to the case of dipole excitation, feedback cooling acts to reduce the amplitude of the quadrupole mode as seen in the reduction of phonon occupancy [Fig.~\ref{fg:quadrupole-results}(a)], although the reduction in energy is not as pronounced as for the dipole modes and is a consequence of the non-linear filtering used to enhance the dipole mode SNR.  When feedback is applied, the temperature is the same as in Fig.~\ref{fg:dipole-results}(f), implying that feedback neither heats nor cools the sample.  Coupling between the quadrupole mode and the thermal gas is weak, resulting in most of the mode's energy dissipating into higher-energy collective modes rather than into the thermal gas.  Instead, we see that the absence of feedback after driving the quadrupole mode leads to cooling of the sample, but this is evaporative cooling driven by the inadvertently-excited vertical dipole mode.  Figure \ref{fg:quadrupole-results}(c) shows that the number of atoms in the sample is lower when feedback is not engaged compared to when it is engaged, and the calculated evaporation efficiency $\gamma = \ln(T'/T)/\ln(N/N') \approx 1.6$ is comparable with independent measurements of the evaporation efficiency for our system at these temperatures.  Evaporative cooling of the sample is not present in Fig.~\ref{fg:dipole-results} as the vertical dipole mode was excited by a sudden increase in optical trap power, leading to a higher trap depth and thus a significantly lower evaporation rate.

\emph{Conclusion and outlook.} Multimode feedback cooling of low-energy modes in a degenerate quantum gas is an important step towards more general feedback cooling schemes.  We have demonstrated feedback cooling of the three lowest-energy collective modes of a BEC, with final energies below one phonon for the dipole modes.  In theory, this is the simplest bottom-up cooling scheme that is capable of cooling a BEC through energy exchange between low- and high-energy modes \cite{haine_control_2004}.  Our feedback setup has the advantage that it can be readily applied to existing experiments, since the control is based on modifications to existing dipole traps, and shadowgraph imaging can be implemented by de-focusing existing imaging systems.  However, our setup is not easily extensible to higher-order modes, so other control architectures such as painted potentials \cite{henderson_experimental_2009} or digital micro-mirror devices \cite{gauthier_chapter_2021} may be needed to achieve the required spatial complexities.  Additionally, a different non-destructive imaging technique, such as phase contrast or Faraday imaging, may be preferable due to the tendency of shadowgraph imaging to amplify low-frequency noise.

Top-down approaches to feedback cooling have challenging requirements, including feedback loops that acquire images, process data, and change external potentials on $\mathord{\sim}\SI{10}{\micro\second}$ timescales \cite{mehdi_fundamental_2024}, with camera-sensor-readout speeds alone needing to be $\mathord{>}\SI{10}{\giga\byte/\second}$.  Further theoretical and experimental work will likely relax these requirements, and technological developments will increase potential feedback-loop response and reduce the cost of such solutions.  In the meantime, bottom-up approaches are more promising avenues for feedback cooling quantum gases.  

\emph{Acknowledgments.} J.A.M acknowledges the efforts and insights of Ryan Husband and Yosri Ben-A\"{i}cha in training him. We acknowledge the Ngunnawal and Ngambri peoples as the traditional custodians of the land on which this research was conducted. This research was funded by Australian Research Council Project No. DP190101709. S.A.H acknowledges support through an Australian Research Council Future Fellowship, Grant No. FT210100809. S.S.S was supported by an Australian Research Council Discovery Early Career Researcher Award, Project No. DE200100495.  R.J.T and S.L. acknowledge support through an Australian Research Council Linkage Project, Grant No. LP19010062.

\bibliographystyle{custom-bib-style}

\end{document}

\title{Supplementary information for: Multi-mode feedback cooling of the collective modes of a Bose-Einstein condensate}
\author{Ryan J. Thomas}\email{ryan.thomas@anu.edu.au}
\author{Jordan A. McMahon}\altaffiliation[Current address: ]{Q-CTRL, 93 Shepherd St, Chippendale, NSW 2008, Australia.}
\author{Zain Mehdi}
\author{Stuart S. Szigeti}
\author{Simon A. Haine}
\author{Samuel Legge}
\author{John D. Close}
\author{Joseph J. Hope}
\affiliation{Department of Quantum Science, Research School of Physics, The Australian National University, Canberra 2601, Australia}

\maketitle

\section{Optical dipole trap configuration and feedback gain calibration}

The atomic sample is trapped at the intersection of laser beams of wavelengths \SI{1064}{\nano\meter} and \SI{1090}{\nano\meter} with beam waists of $w_{\rm 64} = \SI{100}{\micro\meter}$ and $w_{\rm 90} = \SI{150}{\micro\meter}$, respectively.  In a small region around their intersection, the trapping potential is approximately harmonic, and using non-destructive imaging we directly measure the trapping frequencies in the $x$ and $z$ directions to be $f_x = \SI{20.3(6)}{\hertz}$ and $f_z = 70.3(1.3)\,\rm Hz$ by exciting the dipole modes in their respective directions.  We infer the trapping frequency in the $y$ direction to be $f_y = \SI{85.6(6)}{\hertz}$ based on the trap geometry and beam powers.

We implement feedback control of the sample by changing the position at which the \SI{1090}{\nano\meter} laser intersects the \SI{1064}{\nano\meter} laser through piezo-electric actuators on the final steering mirror of the \SI{1090}{\nano\meter} laser (see Fig. 1a in the main text).  We further control the total power in both laser beams by modulating the current in the laser drivers using a voltage-controlled current source.  We represent the state of the actuators by
\begin{equation}
	\mathbf{u} = \begin{bmatrix} V_x\\V_z\\V_{64}\\V_{90}\end{bmatrix},
\end{equation}
for piezo voltages $V_x$ and $V_z$ and power-setting voltages $V_{64}$ and $V_{90}$.  We represent the change to the measured harmonic trap properties by
\begin{equation}
	\mathbf{s} = \begin{bmatrix} \Delta x_{\rm trap} \\ \Delta z_{\rm trap} \\\Delta \omega_x^2\end{bmatrix},
\end{equation}
where $\Delta x_{\rm trap}$ and $\Delta z_{\rm trap}$ are the changes in the trap center positions in the $x$ and $z$ directions, and $\omega_x = 2\pi f_x$ is the angular frequency in the $x$ direction.  The actuator also changes the trap frequencies in the $y$ and $z$ directions, but since we do not control the BEC widths in those directions, we do not include them in our model.  The total open-loop transfer function $\mathrm{G}$, defined by $\mathbf{s} = \mathrm{G}\mathbf{u}$, is then nominally
\begin{equation}
	\mathrm{G} = \begin{bmatrix}
		\SI{-14.4}{\micro\meter/\volt}	& 0	& 0	&	0\\
		0	&	\SI{-1.83}{\micro\meter/\volt}	&	\SI{33}{\micro\meter/\volt}	&	\SI{77}{\micro\meter/\volt} \\
		0	&	0	&	(2\pi\times \SI{26.7}{\hertz})^2/\si{\volt} & (2\pi\times \SI{19.9}{\hertz})^2/\si{\volt}
	\end{bmatrix},
\end{equation}
where some drift in the values occurs over time due to hysteresis in the piezo actuators and changes in electronic voltage offsets.

Our feedback law can be written in terms of a feedback matrix $\mathrm{K}$ as
\begin{equation}
	\mathbf{u}(t_i) = \mathrm{K}[\mathbf{m}(t_i) - \mathbf{m}(t_{i-1})]
\end{equation}
for measurement vector $\mathbf{m}(t_i)$ at time $t_i$, and where the finite difference is approximately $\mathbf{m}(t_i) - \mathbf{m}(t_{i-1}) = \tau \dot{\mathbf{m}}(t_i)$ for time step $\tau = t_i - t_{i-1}$.  The measurement vector is defined by
\begin{equation}
	\mathbf{m} = \begin{bmatrix}
		\hat{x}\\\hat{z}\\\hat{w}_x
	\end{bmatrix},
\end{equation}
for \textit{in situ} estimates of the horizontal and vertical positions $\hat{x}$ and $\hat{z}$, and the estimate of the BEC width $\hat{w}_x$ in the $x$-direction; note that $\hat{w}_x$ is approximately a factor of $1/\sqrt{6}$ smaller than the actual width due to the non-linear filter.  The values for $\mathrm{K}$ are
\begin{equation}
	\mathrm{K} = \begin{bmatrix}
		\SI{-0.82}{\volt/\micro\meter}	&	0	&	0\\
		0	&	\SI{-0.27}{\volt/\micro\meter}	&	0\\
		0	&	0	&	\SI{-0.38}{\volt/\micro\meter}\\
		0	&	0	&	\SI{0.16}{\volt/\micro\meter}	
	\end{bmatrix},
\end{equation}
and the total loop gain, $\mathrm{L} = \mathrm{G}\mathrm{K}$, is nominally
\begin{equation}
	\mathrm{L} = \begin{bmatrix}
		11.8	&	0		&	0\\
		0		&	0.49	&	-0.22\\
		0		&	0		&	-(2\pi\times \SI{14.4}{\hertz})^2/\si{\micro\meter}
	\end{bmatrix}.
\end{equation}
Ideally, $\mathrm{L}$ is diagonal with $\mathrm{L}_{23}$, which represents cross-coupling between control of the horizontal width and the vertical trap position, being exactly $0$.  Imperfect calibration, as well as aforementioned drifts in the open-loop transfer function $\mathrm{G}$, mean that there is almost always some residual gain $\mathrm{L}_{23}$.  Nevertheless, the near-diagonal form of $\mathrm{L}$ significantly reduces cross-coupling of the measurement noise between the modes.

\section{Shadowgraph imaging}

The interaction of a laser beam with electric field $E(\vr)$ and an atomic sample with 3D density $\rho(\vr)$ and on-resonance absorption cross-section $\sigma_0$ can be described using the paraxial wave equation
\begin{equation}
    \left(\partial_y - \frac{i}{2k}\nabla_{\text{T}}^2\right)E(\vr) = -i\frac{\rho(\vr)\sigma_0}{2\tilde{\delta}}E(\vr),
	\label{eq:paraxial-wave-equation}
\end{equation}
for wavenumber $k$ along the $y$ direction, normalized detuning $\tilde{\delta}$ (detuning in half-linewidths), and transverse Laplacian $\nabla_{\text{T}}^2 = \partial_x^2 + \partial_z^2$.  For a thin sample, where we can neglect the diffraction of the laser beam as it passes through the sample, the field $E_f(x,z)$ immediately after passing through the atoms is
\begin{equation}
    E_f(x,z) = E_0(x,z)e^{-i\phi(x,z)},
\end{equation}
for incident electric field $E_0(x,z)$ and phase
\begin{equation}
	\phi(x,z) = \int_{-\infty}^\infty \text{d}y \frac{\rho(\vr)\sigma_0}{2\tilde{\delta}} = \frac{\OD(x,z)}{2\tilde{\delta}},
\end{equation}
for on-resonance optical depth $\OD(x,z)$.  After the field has passed through the atoms, the formal solution to Eq.~\eqref{eq:paraxial-wave-equation} after propagation by a distance $\xi$ along the $y$-axis is
\begin{align}
	E(x,z;\xi) &= \exp\left(\frac{i\xi}{2k}\nabla_{\text{T}}^2\right)E_f(x,z)\\
	&\simeq \left(1 + \frac{i\xi}{2k}\nabla_{\text{T}}^2\right)E_f(x,z)
\end{align}
where the approximation is that the distance $\xi$ is small enough that the factor $(\xi/k)\nabla_{\text{T}}^2E_f(x,z) \ll 1$.  If the incident field is uniform over the spatial extent of $\phi(x,z)$, then we can write the intensity $I = 2\epsilon_0 c |E|^2$ as
\begin{equation}
	I(x,z;\xi) \approx I_0(x,z)\left(1 - \frac{\xi}{k}\nabla_{\text{T}}^2\phi(x,z)\right),
	\label{eq:shadowgraph-intensity}
\end{equation}
and the phase written to the light by the atoms now appears as a change in the measured intensity.  In our experiment, $\xi$ is the distance between the atoms and the imaging system's object plane.  For the \textit{in situ} data processing used for feedback, we invert Eq.~\eqref{eq:shadowgraph-intensity} to infer the 2D phase, which is proportional to density.  We can write the phase as
\begin{align}
	\phi(x,z) &\propto \nabla_{\mathrm{T}}^{-2}\left[\frac{I(x,z;\xi)}{I_0(x,z)} - 1\right]\\
    &= \mathcal{F}^{-1}\left[\frac{1}{k_x^2 + k_z^2}\mathcal{F}\left[\frac{I(x,z;\xi)}{I_0(x,z)} - 1\right]\right],
	\label{eq:fft-inverse-laplacian}
\end{align}
where $\mathcal{F}[\cdot]$ is the Fourier transform and $\mathcal{F}^{-1}[\cdot]$ is its inverse.  This Fourier-based inverse is efficiently computed in real time.  To regularize the singularity at the origin in the Fourier domain we set the measurement current, $I(x,z;\xi)/I_0(x,z) - 1$, to zero, and then adjust the offset of the inferred $\phi(x,z)$ so that it is zero in the region where the atomic density is known to always be zero. 

A second complication is that the incident intensity must be known.  We measure a reference image for each run of the experiment by taking a shadowgraph image during one of our early evaporative cooling steps.  At this stage, the atomic ensemble is thermal with a temperature on the order of $\mathord{\sim}\SI{10}{\micro\kelvin}$, so that the spatial extent is much larger than that of the BEC and the optical depth is very low, leading to a negligible change in the intensity of the beam.  This becomes our reference image, $I_0(x,z)$.  

In the main text, we describe fitting a model of the shadowgraph images to the raw data, and we do this for two main reasons.  First, feedback stabilizes the measurement $\mathbf{m}$ comprised of the \textit{in situ} estimates, but that does not necessarily mean that feedback stabilizes the system state due to measurement noise.  By using a different method of estimating the mode amplitudes, with independent measurement noise, we can faithfully infer the performance of feedback.  Second, the \textit{in situ} estimates prioritize processing speed over noise reduction, while for offline characterization we can use a computationally intensive 2D fit to reduce measurement noise.

Our shadowgraph model computes the expected intensity by propagating the paraxial wave equation in the Fourier domain to a propagation distance $\xi$.  We assume that the measured phase is dominated by the condensed component of the atomic sample, so that the phase has a spatial profile of
\begin{equation}
	\phi(x,z) = \begin{cases}
		\phi_0\left(1 - \frac{(x - x_0)^2}{R_x^2} - \frac{(z - z_0)^2}{R_z^2}\right)^{3/2}, & \frac{(x - x_0)^2}{R_x^2} + \frac{(z - z_0)^2}{R_z^2} \leq 1\\
		0, & \frac{(x - x_0)^2}{R_x^2} + \frac{(z - z_0)^2}{R_z^2} > 1
	\end{cases}
\end{equation}
for Thomas-Fermi radii $R_{x,z}$ and cloud center $(x_0,z_0)$.  The electric field at the object plane is then
\begin{equation}
    E(x,z;\xi) \propto \mathcal{F}^{-1}\bigg[\exp\left({-\eta^2(k_x^2 + k_z^2)}\right)\exp\left({\frac{i \xi}{2k}(k_x^2 + k_z^2)}\right)\mathcal{F}\bigg[\exp\left({-i\phi(x,z)}\right)\bigg]\bigg]
	\label{eq:shadowgraph-model}
\end{equation}
where $\eta \approx \SI{5.5}{\micro\meter}$ is the resolution of the imaging system.  The intensity at the camera is then $I(x,z) \propto |E(x,z)|^2$.  We calibrate $\xi$ by fitting Eq.~\eqref{eq:shadowgraph-model} to a set of images where the sample is not perturbed, and we find that $\xi\approx \SI{800}{\micro\meter}$, which agrees well with numerical calculations that indicate that the maximum of the shadowgraph signal occurs when $\xi$ is between \SI{500}{\micro\meter} and \SI{1}{\milli\meter}.

\section{\textit{In situ} feedback calculation}

As noted above, for each run of the experiment we collect a reference image $I_0(x,z)$, and then we measure the current intensity $I_n(x,z)$ for each image.  We use the fast Fourier transform (FFT) based filter in Eq.~\eqref{eq:fft-inverse-laplacian} to get a scaled estimate of the column density for the $n$-th image
\begin{equation}
	\hat{\rho}_n(x,z) \propto \mathcal{F}^{-1}\left[\frac{1}{k_x^2 + k_z^2}\mathcal{F}\left[\frac{I_n(x,z)}{I_0(x,z)} - 1\right]\right],
\end{equation}
where, as described previously, we define the point at $(k_x,k_z) = (0,0)$ to be $0$.  We then subtract from $\hat{\rho}_n(x,z)$ the mean value of a region of pixels where there is no atomic density so that the estimated atomic density is zero outside of the region where we expect the BEC.  

We calculate the $n$-th mean position of the cloud and the width in the $x$ direction through moments
\begin{equation}
	\expect{x^k_n} = \frac{\sum_z\sum_x \hat{\rho}_n(x,z) x^k}{\sum_z\sum_x \hat{\rho}_n(x,z)},
	\label{eq:moment-no-filter}
\end{equation}
and similarly for $z$, as these are the most computationally efficient methods for estimating the mode amplitudes.  We exclude from the summations regions of the image where the atomic density is always $0$.  For this moment method to work, however, $\hat{\rho}_n(x,z) \geq 0$, and due to measurement noise and limitations of Eq.~\eqref{eq:shadowgraph-intensity} we can have $\hat{\rho}(x,z) < 0$.  Furthermore, moment calculations are sensitive to outliers, and the low frequency noise generated by the inverse Laplacian filter significantly lowers the SNR of our estimates.

To address this problem, we use a non-linear filter, $[\hat{\rho}_n(x,z)]^6$, to improve the SNR of the moment calculations.  In regions of $\hat{\rho}$ where its SNR is $\mathord{>}1$, which is in the region where the density is high, we see an enhancement of the image SNR.  We find that this improves the SNR for the moment calculations, which are now calculated as
\begin{equation}
	\expect{x^k_n} = \frac{\sum_z\sum_x [\hat{\rho}_n(x,z)]^6 x^k}{\sum_z\sum_x [\hat{\rho}_n(x,z)]^6},
	\label{eq:moment-filter}
\end{equation}
where, again, the summation is over a restricted region.  In the absence of measurement noise, the mean values $\expect{x}$ and $\expect{z}$ calculated using Eq.~\eqref{eq:moment-filter} will be the same as those calculated using Eq.~\eqref{eq:moment-no-filter}, while the variance $\expect{x^2} - \expect{x}^2$ will be smaller when using the nonlinear filter ($1/6$ as large for a Gaussian distribution).  Changes in the scale factors are then absorbed into the definition of the feedback gains, and since we are only concerned with derivative control, offsets in the estimates do not affect feedback.

\section{Calculation of phonon occupancy}

The Hamiltonian describing the energy of the $i$-th mode of the system, denoted $r_i$, is 
\begin{equation}
	H_i = \frac{1}{2}m\omega_i^2 (r_i - r_{i,\rm trap})^2 + \frac{1}{2}m \dot{r}_i^2
\end{equation}
where $r_{i,\rm trap}$ is the equilibrium position of the trapping potential.  For the moment, let us assume that $r_{i,\rm trap} = 0$; then, the average energy $\expect{E}$ is
\begin{align}
	\expect{E} = \left(n_i + \frac{1}{2}\right)\hbar\omega_i &= \frac{1}{2}m\omega_i^2\expect{r_i^2} + \frac{1}{2}m\expect{\dot{r}_i^2}\\
	&= \frac{1}{2}m\omega_i^2\bar{r}_i^2(t) + \frac{1}{2}m\dot{\bar{r}}_i^2(t) + \frac{1}{2}m\omega_i^2\var(r_i) + \frac{1}{2}m\var(\dot{r}_i),
\end{align}
where the average is over the atomic ensemble.  We write the mean position of the sample and its velocity as $\bar{r}_i(t)$ and $\dot{\bar{r}}_i(t)$, respectively.  The terms related to the variance in $r_i$ and $\dot{r}_i$ contribute the zero-point energy $\frac{1}{2}\hbar\omega_i$, and, as is well-established in the optomechanics literature, we will ignore this contribution in our calculations of average phonon number $n_i$.  The average phonon number is then
\begin{equation}
	n_i = \frac{m\omega_i^2\bar{r}_i^2(t) + m\dot{\bar{r}}_i^2(t)}{2\hbar\omega_i} = \frac{\bar{r}_i^2(t)}{2a_{{\rm ho},i}^2} + \frac{\dot{\bar{r}}_i^2(t)}{2a_{{\rm ho},i}^2\omega_i^2} 
	\label{eq:phonon-number-1}
\end{equation}
where $a_{{\rm ho},i} = \sqrt{\hbar/(m\omega_i)}$ is the harmonic oscillator length for mode $i$.  Although in the absence of feedback $n_i$ is independent of time, with feedback present $n_i$ will be time-dependent.

To ensure consistency in measuring $n_i$ regardless of whether feedback is disabled, so that $\bar{r}_i(t)$ undergoes simple harmonic motion, or when feedback is enabled, so that $\bar{r}_i(t)$ is ideally uncorrelated white noise, we calculate the phonon number $n_i$ as
\begin{equation}
	n_i = \frac{1}{2}m\omega_i^2\var_t\big[\bar{r}_i(t) - r_{i,\rm trap}(t)\big] + \frac{1}{2}m\var_t\big[\dot{\bar{r}}_i(t)\big]
\end{equation}
where $\var_t(\cdot)$ is calculated using the measurement record over the last oscillation period.  We include the trap position for the dipole modes only.

Noise in measuring $\bar{r}_i(t)$ biases the measurement of energy.  Denoting the measurement noises as $\delta\bar{r}_i$ and $\delta\dot{\bar{r}}_i$, and replacing $\bar{r}_i\to \bar{r}_i + \delta\bar{r}_i$ and similarly for $\dot{\bar{r}}_i$, the measured phonon number is
\begin{align}
	n_{i,\rm meas} &= \underbrace{\Big(\frac{1}{2}m\omega_i^2\var_t\big[\bar{r}_i(t) - r_{i,\rm trap}(t)\big] + \frac{1}{2}m\var_t\big[\dot{\bar{r}}_i(t)\big]\Big)}_{n_{i,\rm true}} + \frac{1}{2}m\big[\omega_i^2\var(\delta\bar{r}_i) + \var(\delta\dot{\bar{r}}_i)\big],
\end{align}
where $n_{i,\rm true}$ is the true phonon occupancy.  It is important to note that $n_{i,\rm true}$ includes the effect of \textit{in situ} measurement noise, whereas $n_{i,\rm meas}$ also includes the bias from measurement noise in post-processing.  $\delta\bar{r}_i$ is uncorrelated white noise, and we calculate $\dot{\bar{r}}_i(t_j) = [\bar{r}_i(t_j) - \bar{r}_i(t_{j-1})]/\tau$ for time step $\tau$ so that $\var(\delta\dot{\bar{r}}_i) = 2\var(\delta\bar{r}_i)/\tau^2$.  Therefore, the true phonon occupancy is
\begin{equation}
	n_{i,\rm true} = n_{i,\rm meas} - \frac{1}{2}\omega_i^2\var(\delta\bar{r}_i)\left[1 + \frac{2}{\omega_i^2\tau^2}\right].
\end{equation}

\section{Comparison with out-of-trap measurements}

To compare with out-of-trap measurements, we assume that when feedback is turned off the energy during the hold time is the same as the energy at the end of feedback.  Furthermore, when feedback is off, the ensemble undergoes simple harmonic motion, so the energy is, on average, evenly distributed between the position and velocity degrees of freedom.  Using the horizontal degree of freedom as an example, the horizontal position of the ensemble after a time-of-flight $\tof$ is
\begin{equation}
	x(\tof) = x + v\tof,
\end{equation}
where $x$ and $v$ here are the position and velocity of the ensemble when the trap is switched off.  The variance in $x(\tof)$ is then
\begin{align}
	\var(x(\tof)) &= \var(x) + \var(v)\tof^2\\
	&= \frac{E}{m\omega_x^2} + \frac{E}{m}\tof^2\\
	&= \frac{E}{m\omega_x^2}(1 + \omega_x^2\tof^2).
\end{align}
A similar result holds for the vertical degree of freedom.